\documentclass[aps,twocolumn,prd,showpacs,preprintnumbers,nofootinbib,amsmath,amssymb]{revtex4}

\usepackage{graphicx}
\usepackage{dcolumn}
\usepackage{bm}
\usepackage{amsmath}

\newcommand{\beq}{\begin{eqnarray}}
\newcommand{\eeq}{\end{eqnarray}}

\newcommand{\real}{{\sf I}\kern-.12em{\sf R}}
\newcommand{\comp}{{\sf I}\kern-.50em{\sf C}}
\newcommand{\unity}{{\sf I}\kern-.54em{\sf 1}}

\def\spose#1{\hbox to 0pt{#1\hss}}
\def\ltapprox{\mathrel{\spose{\lower 3pt\hbox{$\mathchar"218$}}
 \raise 2.0pt\hbox{$\mathchar"13C$}}}

\begin{document}


\title{The order of the Roberge-Weiss endpoint (finite size transition)
in QCD.}
\author{Massimo D'Elia$^{1}$ and Francesco Sanfilippo$^{2}$}
\affiliation{$^1$Dip. di Fisica, Universit\`a
di Genova and INFN, Via Dodecaneso 33, 16146 Genova, Italy\\
$^2$Dip. di Fisica, 
Universit\`a di Roma ``La Sapienza'' and INFN, P.le A. Moro 5, 00185 Roma, Italy}

\date{\today}

\begin{abstract}
We consider the endpoint of the Roberge-Weiss (RW) first order transition line present
for imaginary baryon chemical potentials. We remark that it coincides with the 
finite size transition relevant in the context of large $N_c$ QCD and
study its order in the theory with two degenerate flavors.
The RW endpoint is first order in the limit of large and small
quark masses, while it weakens for intermediate masses where it is likely in the 
Ising 3d universality class. Phenomenological implications and further
speculations about the QCD phase diagram are discussed.
\end{abstract}

\maketitle

\section{Introduction}

The determination of the QCD phase diagram at finite temperature T and baryon chemical $\mu_B$
is one of the outmost 
open problems within
the Standard Model of Particle Physics. At $\mu_B = 0$, lattice QCD simulations
have shown the presence of a finite T deconfinement/chiral symmetry restoring transition:
in the limit of zero or infinite quark masses it is associated to a change in the 
realization of some exact symmetry (chiral or center symmetry respectively), hence it must be a 
real phase transition. For finite quark masses no further exact symmetries are presently known, 
hence the transition may also be a smooth crossover.
At $\mu_B \neq 0$ the deconfinement line starting from  
$\mu_B = 0$ may merge with an analogous transition line starting from the $T = 0$
axis: the latter could be first order and a critical endpoint may be present if the $\mu_B = 0$ transition is a crossover. Large efforts are dedicated to the theoretical
and experimental search of this possible endpoint: unfortunately lattice QCD 
simulations at nonzero $\mu_B$ are hindered by the sign problem and a direct numerical
investigation is hardly feasible.

A way to partially overcome the sign problem is to consider an
imaginary chemical potential, $\mu_B = i \mu_{I}$:
numerical simulations are feasible 
and information about real $\mu_B$ can be recovered by 
analytic continuation techniques~\cite{alford,lombardo,muim,muim2,immu_dl,azcoiti,chen,giudice,cea,sqgp,conradi,cea2,sanfo1,cea3}. 
Definite answers can be obtained regarding the structure of the phase diagram in the 
$T$-$\mu_I$ plane:
those answers do not give direct information on the finite density phase diagram but, 
as remarked in recent literature~\cite{sqgp,Kouno:2009bm} 
and as further stressed in the present study, 
they may be relevant to physics at zero or small real $\mu_B$. 

One of the main features of the $T$-$\mu_I$ diagram are the first order lines
met at high $T$ and fixed periodic values of $\mu_I$
known as Roberge-Weiss (RW) transitions~\cite{rw}.
The end of such lines is a phase 
transition itself when moving in $T$ at fixed $\mu_I$:
its order may be relevant to QCD phenomenology
and is the subject of our study.

The $\mu_B =0$ partition function 
in the lattice formulation is
\beq
Z(T) \equiv \int \mathcal{D}U e^{-S_{G}[U]} \det M[U]
\eeq
where $U$ stands for gauge link variables, $S_G$ is the pure gauge action and $M$ is the fermionic matrix:
periodic (antiperiodic) boundary conditions are understood for boson (fermionic) fields in the Euclidean 
time direction. More fermion determinants or powers of them may be needed depending 
on the flavor spectrum.

In the quenched limit (no fermion determinant) 
the action is symmetric
under
multiplication of all temporal links at a given time by an element
of the center of the gauge group 
$Z_{N_c}\equiv\left\{e^{i 2 k \pi/N_c}, k = 0, \dots  N_c - 1 \right\}$.
This is known as center symmetry and
gets spontaneously broken at the deconfinement 
transition. The Polyakov loop $L$
is not invariant and serves as an exact order parameter: $\langle L \rangle$ is nonzero and proportional
to a center element in the deconfined phase. 
The fermion determinant breaks center
symmetry explicitely: $\langle L \rangle $ is always nonzero and real,
but it still rapidly increases at the transition.

In order to discuss the phase diagram in $T$-$\mu_I$ plane,
we introduce the dimensionless variable $\theta_q \equiv {\rm Im} (\mu_q)/T$, where
$\mu_q = \mu_B/3$ is the quark chemical potential: $\mu_I$
can be interpreted as a constant $U(1)$ background field in the Euclidean 
time direction or, equivalently, 
as a twist in the fermionic temporal boundary conditions by an angle $\theta_q$,
\beq
Z(T,\theta_q) \equiv \int \mathcal{D}U e^{-S_{G}[U]} \det M[U,\theta_q]
\eeq
A transformation $\theta_q \to \theta_q + 2 \pi k/N_c$
can be exactly cancelled by a center transformation leaving both
$S_G$ and the functional integration invariant,
hence the free energy is periodic in $\theta_q$ with 
period $2 \pi/N_c$~\cite{rw}, instead of the expected $2 \pi$.
Such periodicity is smoothly realized at low $T$,
while at high $T$ 
phase transitions occur for $\theta_q = (2 k + 1)\pi/N_c$ and $k$ integer,
at which $\langle L \rangle$ jumps from one center sector to the other: the phase of $L$ and
other physical observables (e.g. the baryon density) are discontinuous at these points~\cite{rw}.
The emerging picture for the $T$-$\theta_q$ 
phase diagram is that of a periodic repetition of first order
lines (RW lines) in the high $T$ regime, which
must disappear in the low $T$ regime, hence they have an endpoint at some temperature
$T_{\rm RW}$. 
There is numerical evidence that the 
analytic continuation to imaginary $\mu_B$ 
of the deconfining/chiral restoring line (present for real $\mu_B$) touches
the RW line right on its endpoint and then repeats periodically in the $T$-$\theta_q$ plane:
we shall comment on this issue later. 
The deconfining temperature increases with $\mu_I$,
hence $T_{\rm RW} > T_c$, where $T_c$ is the critical temperature at $\mu_B = 0$.

RW lines corresponds to points where a $Z_2$ symmetry is spontaneously broken:
two center sectors are equivalent but one of them is selected.
Let us consider in particular the 
$\theta_q = \pi$ line. As for $\theta_q = 0$, 
there is symmetry with respect to
complex conjugation of link variables, i.e. charge conjugation, however in the high $T$
phase for $\theta_q = \pi$ the system selects one of two center sectors in which $\langle \L \rangle$
is complex and charge conjugation is spontaneously broken: $T_{\rm RW}$
is the critical temperature at which charge symmetry breaks, hence it corresponds 
to a real phase transition independently of the quark spectrum. 
The RW line is analogous to the first order line lying along the $T$ axis 
of an Ising system: the endpoint plays the role of the Ising critical $T$ 
and $(\theta_q - \pi)$ that of
the magnetic field.
It is worth stressing that instead at $\theta_q = 0$
$\langle \L \rangle$ is always real and charge conjugation stays unbroken.

The RW endpoint can be given a different interpretation. Setting $\theta_q = \pi$ is like 
switching from antiperiodic to periodic boundary conditions for fermion fields in the
Euclidean time direction. 
The system can be viewed 
as a usual thermal system in presence of an imaginary chemical             
potential or, naming differently the axes, as a $T = 0$ system with one spatial dimension compactified:
in the second case the RW endpoint defines a critical size $L_c = 1/T_{\rm RW}$ 
of the compact dimension, below which charge conjugation is spontaneously broken.

Such transition has been studied in recent literature~\cite{degrand0,degrand,lucini,lucini2} and is 
relevant to large $N_c$ QCD, in particular to
large $N_c$ orientifold planar equivalence~\cite{asv}. That states that 
QCD with fermions in the antisymmetric representation, QCD(AS), which coincides
with ordinary QCD for $N_c = 3$, is equivalent, 
in the large $N_c$ limit and in the charge-even sector of the theory,
to QCD with fermions in the adjoint representation. 
Such equivalence is guaranteed if the charge conjugation symmetry
is not spontaneously
broken in QCD(AS)~\cite{uy2006} (actually the condition
may be not so strict, as discussed in Ref.~\cite{myers}).
That 
fails in presence of a compact dimension, below a critical
compactification radius, which is nothing but a different mapping of 
the endpoint temperature $T_{\rm RW}$.

The existence of a phase transition at the RW endpoint is fixed by symmetry, but its 
nature is not.
If it is second order, 
symmetry suggests that
it belongs to the universality class of the 3d Ising model; the
corresponding critical behaviour could in principle have influence also far from the endpoint, 
e.g. for zero density QCD right above $T_c$, 
as suggested by recent literature~\cite{sqgp,Kouno:2009bm}. 
However it 
may also be first order: this is the case in the infinite quark mass limit, where the transition
coincides with the pure gauge SU(3) thermal transition.
In general the answer may depend on the flavor spectrum.

In the first order case, the RW endpoint is actually a triple point, with
two further first order lines departing from it for non-zero
values of the magnetic variable (think of the $3d$ 3-state Potts model as a similar example). 
It would be 
natural to identify this departing line with (part of) the analytic continuation to imaginary $\mu_B$ of the
physical line, thus explaining why it meets the RW line right on its endpoint.
The departing line could reach the $\mu_I = 0$ axis or 
have a second order critical endpoint arbitrarily close
to it, thus with a strong influence on the properties of strong interactions right above
$T_c$.

In the following we show results regarding QCD with two degenerate flavors. Earlier lattice results~\cite{muim}
on small lattices suggested a second order nature, supported recently by the analysis
of effective models~\cite{Kouno:2009bm}. Instead we provide evidence that the transition
is first order for low enough masses and that it weakens
for intermediate quark masses, where it could be second order.

\begin{figure*}[t!]
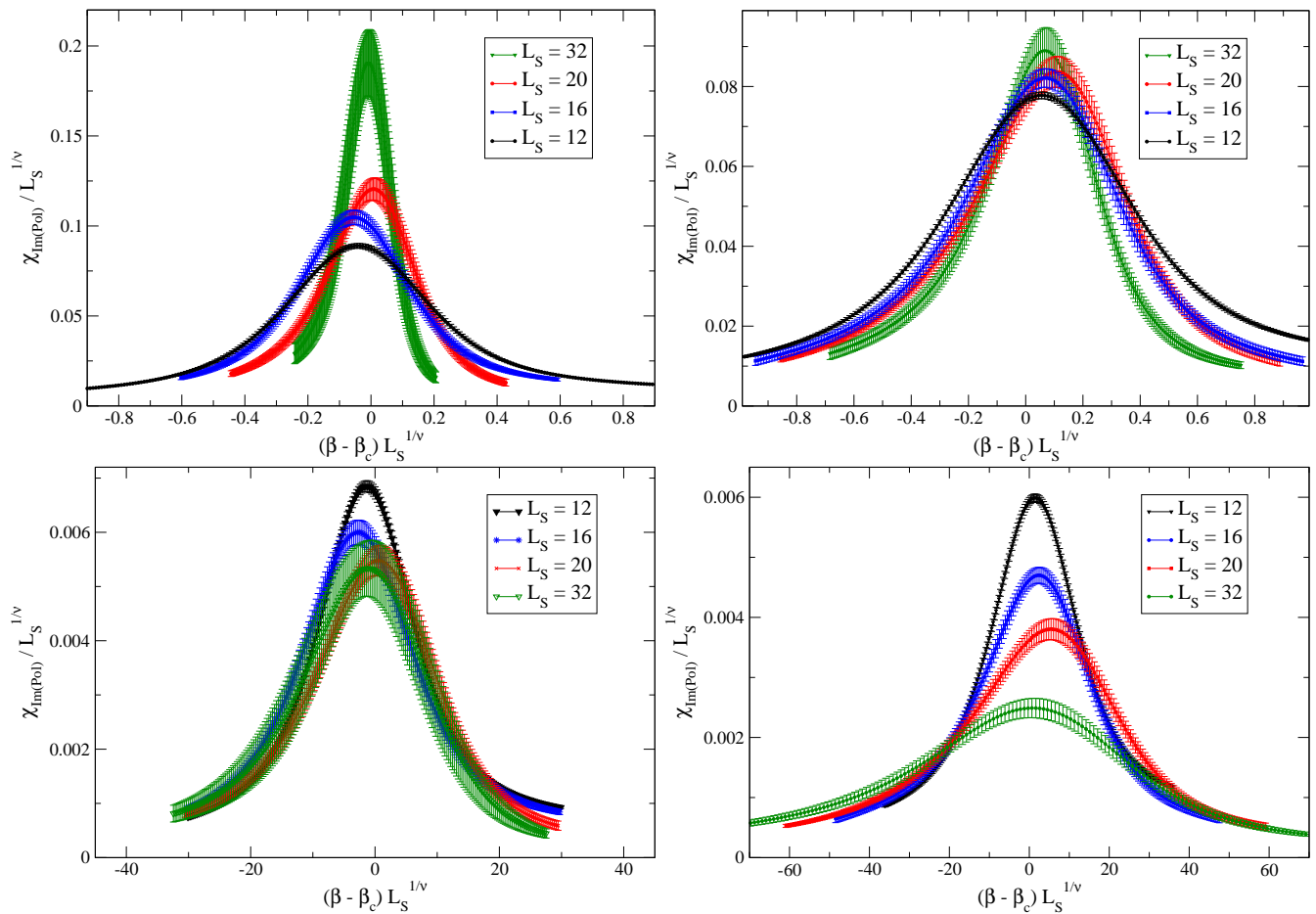

\includegraphics*[width=0.49\textwidth]{scaling_0.025_2nd.eps}
\includegraphics*[width=0.49\textwidth]{scaling_0.075_2nd.eps}\\
\includegraphics*[width=0.49\textwidth]{scaling_0.025_1st.eps}
\includegraphics*[width=0.49\textwidth]{scaling_0.075_1st.eps}
\caption{Scaling of the reweighted susceptibility of the imaginary part of the Polyakov loop
for $am = 0.025$ (left column) and $am = 0.075$ (right column) according to the Ising 3d universality class
(upper row) or to first order critical indexes (lower row).}
\label{scaling}
\end{figure*}

\begin{figure*}[t!]
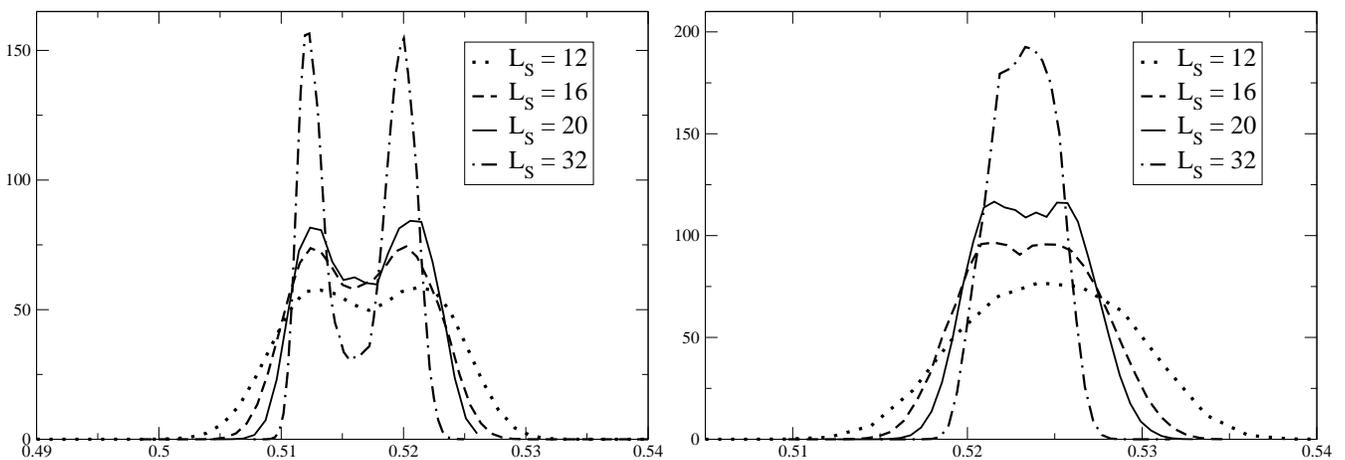

\includegraphics*[width=0.49\textwidth]{isto0.025.eps}
\includegraphics*[width=0.49\textwidth]{isto0.075.eps}
\caption{Reweighted plaquette distribution at $\beta_c$
as a function of $L_s$ for $am = 0.025$ (left) and $am = 0.075$ (right).}
\label{plaqdis}
\end{figure*}

\section{Numerical Results}

We have investigated QCD with two degenerate flavors, adopting the standard 
plaquette action, the standard staggered 
fermion formulation, and using a Rational Hybrid Monte Carlo algorithm.
Two values of the quark mass have been explored, $a m_q = 0.075$ and
$a m_q = 0.025$, the latter coinciding with that used in Ref.~\cite{muim}.
In order to perform a finite size scaling analysis at the critical
endpoint, we have made simulations
on lattices $L_s^3 \times L_t$ with $L_t = 4$ and 
$L_s = 8,\ 12,\ 16,\ 20,\ 32$. We have worked at fixed $\theta_q = \pi$
and the temperature 
$T = 1/(L_t a(\beta,m_q))$ has been changed by tuning the inverse gauge coupling $\beta$.
Collected statistics
are of the order of $50 - 100$K trajectories for the $\beta$ values closest 
to the critical point.
Simulations have been performed on a computer farm in Genoa, apart
from those on the biggest lattice, $L_s = 32$, for which the
apeNEXT facilities in Rome have been used. Both resources have been provided by INFN.

Since we work at $\theta_q = \pi$, we have chosen the imaginary part
of the Polyakov loop 
as an order parameter: it is not invariant under
charge conjugation and develops a nonzero expectation value for 
$T > T_{\rm RW}$ (another possible choice could have been the
imaginary part of the baryon density~\cite{lucini,Kouno:2009bm}):
actually we consider its modulus for the finite size scaling analysis,
similarly to what is done for the magnetization of an Ising system.
Its susceptibility
\beq
\chi \equiv L_s^3\ (\langle {\rm Im}(L)^2 \rangle - \langle |{\rm Im}(L)| \rangle^2) 
\eeq 
where $L$ is the spatially averaged Polyakov loop trace 
(normalized to $N_c$), 
is expected to scale as follows:
\beq
\chi = L_s^{\gamma/\nu}\ \phi (\tau L_s^{1/\nu}) \, .
\label{fss}
\eeq
That means that 
the quantities $\chi/L_s^{\gamma/\nu}$ measured on different lattice sizes 
should fall on the same curve when plotted against $\tau L_s^{1/\nu}$. 
The critical indexes are $\nu \sim 0.63$ and 
$\gamma \sim 1.24$ for the 3d Ising model,
while a first order transition is effectively described,
in three spatial dimension, by $\nu = 1/3$ and $\gamma = 1$.
In the following $(\beta - \beta_{\rm RW})$ will be used in place of
$\tau \equiv (T - T_{\rm RW})/T_{\rm RW}$: that can be done close enough to the critical point.
In Fig.~\ref{scaling} we show how these two scaling ans\"atze work for 
the two masses explored; the plotted susceptibilities have been obtained by 
Ferrenberg-Swendsen reweighting. At the lowest quark mass the scaling with 
3d Ising critical indexes clearly fails (upper-left figure), while 
first order scaling is good (lower-left figure), in both cases we have used $\beta_{\rm RW} = 5.33885$:
these results are in disagreement with those of Ref.~\cite{muim} and this is 
likely due to the small volumes $L_s = 6,8$ and to the non exact 
molecular dynamics algorithm used in Ref.~\cite{muim}.
 
Instead at the larger quark mass, $a m_q = 0.075$, first order scaling 
is not good (lower-right figure) while 
better, even if not perfect, agreement is found with the 3d Ising critical 
behaviour (upper-right figure). 
In this case the critical coupling has been set to $\beta_{\rm RW} = 5.3965$.

Consistent results are obtained for the plaquette distribution
at the critical couplings, reported in Fig.~\ref{plaqdis}. For $a m_q = 0.025$ 
a double peak structure develops, becoming sharper and sharper
as the volume increases; a similar behaviour (not shown in the figure)
is observed for the chiral condensate, which is also discontinuos at the transition.
This is not the case for $a m_q = 0.075$:
we cannot exclude a very weak first order transition with discontinuities
better visible on volumes larger than those explored by us, but we conclude that at $a m_q = 0.075$
the transition is surely much weaker than at $a m_q = 0.025$.

The simplest scenario compatible with our results is that
the RW endpoint is first order both for very heavy or light quarks,
while it weakens and could be second order for
intermediate masses. 
Note that $a m_q = 0.025$ corresponds to a pion mass already quite larger than the physical
one, therefore we predict a first order
RW endpoint also for physical quark masses. 
Our results should be checked closer to 
the continuum limit using improved actions and/or larger values of $L_t$; also
a check with different fermion formulations would be important. 
Nevertheless, if this scenario is confirmed, it has important phenomenological 
consequences and gives rise to further speculations that we discuss in the next Section.

\section{Discussion and speculations}

\begin{figure}[h!]
\includegraphics*[width=0.48\textwidth]{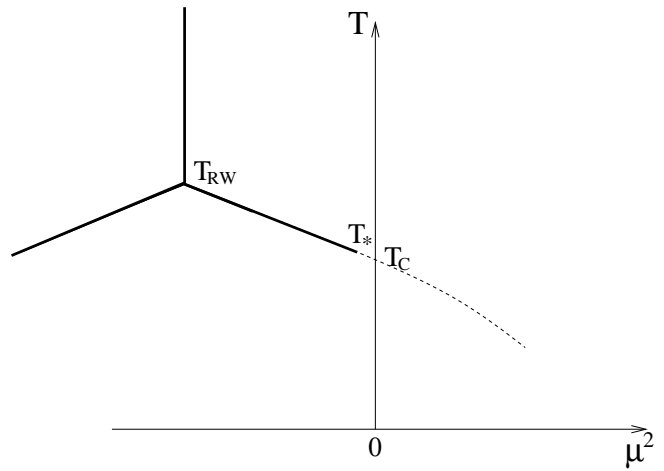}
\caption{A possible sketch of the $T - \mu^2$ diagram in which 
the first order line departing from the RW endpoint ends at a critical
temperature $T_*$ before reaching
the $\mu^2 = 0$ axis.}
\label{pos1}
\end{figure}

A first order RW endpoint implies further
first order lines departing from it.
One of those lines is part of the analytic continuation 
of the physical transition line: 
the fate of this first order line is of great importance.
Two possibilities can be realized:

{ i)}
It ends in a critical endpoint before reaching the $\mu_I = 0$ axis,
as sketched in Fig.~\ref{pos1}. However the endpoint could be close
to the axis and thus have strong influence on zero density
physics slightly above $T_c$;

{ ii)}
It reaches the $\mu_I = 0$ axis and possibly goes through it,
implying a first order transition at zero density and also
for small real chemical potentials.

It is reasonable to assume that the extension of the first order line
increases as the strength of the RW endpoint transition increases. Hence, according to
our results, the second possibility should be more likely for very heavy or very light quarks: 
indeed the possible presence of a first order transition
in the chiral limit of $N_f = 2$ QCD is not excluded by recent lattice 
studies~\cite{2flv1,2flv2}.

The strengthening of the RW endpoint transition for large quark masses is understandable,
since the symmetry which is spontaneously broken at the endpoint is a remnant $Z_2$ subgroup of the 
original center symmetry, which is exact in the quenched limit. Instead the strengthening in 
the chiral limit is an unexpected phenomenon, which is likely linked to the interplay with
chiral degrees of freedom and which should be investigated by future studies. It is reasonable 
to assume that such phenomenon takes place also for different number of flavors: then
further speculations follow.

\begin{figure}[h!]
\includegraphics*[width=0.49\textwidth]{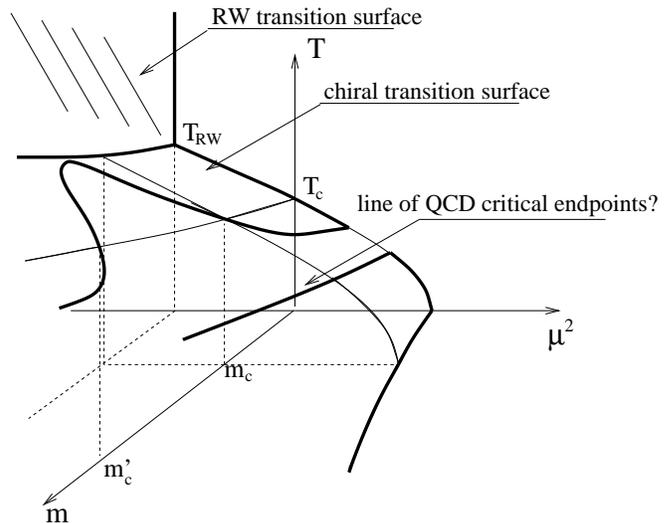}
\caption{Sketch of a speculative phase diagram 
for $N_f = 3$ case, inspired by our
results for $N_f = 2$.
Thick solid lines represent the border of first order transition surfaces.}
\label{figdiag}
\end{figure}

Consider the case of three degenerate flavors,
$N_f = 3$, where it is known 
that, at $\mu_B = 0$, the transition is first
order for small or large quark masses, with two critical quark masses $m_c$ and
$m_c'$ delimiting an intermediate region where a crossover is present (think of the 
Columbia plot). It is possible to interpret the two first order regions in terms of the 
realization of possibility ii) described above: for $m < m_c$ or $m > m_c'$ the RW
endpoint is so strong that the departing first order line reaches the 
$\mu_B = 0$ axis; the same does not happen instead for intermediate quark masses
$m_c < m < m_c'$. This conjecture is sketched in Fig.~\ref{figdiag}, representing
a speculative phase structure for $N_f = 3$ in the $T$-$m$-$\mu^2$ diagram 
(imaginary chemical potentials corresponds to $\mu^2 < 0$): a first order surface
departs from the line of RW endpoints and extends enough to reach $\mu^2 = 0$ only for
low or high $m$.

The diagram in Fig.~\ref{figdiag} is in contrast
with a standard scenario in which the first order region seen for
$\mu^2 = 0$ and $m < m_c$ is the intersection with a first order surface related
to high $\mu^2$, low $T$ physics and to which the critical endpoint of the QCD 
phase diagram belongs, so that the critical endpoint 
is on the same line as the $m = m_c$, $\mu^2 = 0$ point. 
In Fig.~\ref{figdiag} the high $\mu^2$, low $T$ surface is plotted as well
(this is a simplification, since
more transition surfaces may be present at high $\mu^2$), but it is not related
to the first order transition
seen for $\mu^2 = 0$, $m < m_c$, so that
if a QCD critical
endpoint exists, it is not related 
to the chiral critical surface.

Fig.~\ref{figdiag} is a conjecture for $N_f = 3$ inspired by our results 
for the $N_f = 2$ RW endpoint: is it
a reasonable guess? Future numerical studies can 
clarify that, at least for the $\mu^2 < 0$ side. Moreover,
if we consider how $m_c$ changes as $\mu^2$ is slightly increased from
zero, 
then according to Fig.~\ref{figdiag} it should decrease,
while according to the standard scenario it should increase:
the authors of Refs.~\cite{deph1} have provided 
evidence showing that $m_c(\mu^2)$ is indeed a decreasing function of $\mu^2$,
at least for small $\mu^2$, thus supporting a scenario like that
shown in Fig.~\ref{figdiag}.
\\

To summarize, we have provided evidence that the endpoint
of the RW transition for $N_f = 2$ 
is first order for small or high quark masses, while it weakens and could be second
order for intermediate masses. A first order RW endpoint implies
two further first order lines departing from it: it is natural to identify
those lines with part of the analytic continuation of the physical line, thus explaining
why the latter and the RW line are connected to each other.
The range of light quark masses, for which the RW
endpoint is first order, includes physical quark masses, 
it is therefore of great importance to understand 
what is the fate of the further first order line departing from the endpoint: 
it may reach the zero density axis or have a critical endpoint arbitrarily close to it,
which could have great influence on the physics of strongly interacting matter right above the 
deconfinement transition. We will clarify this point in the future,
by studying how fast the first order observed at the RW endpoint weakens 
as $\mu_I$ moves towards the $\mu = 0$ axis.
A careful check of our results closer to the continuum limit should
be performed as well.
Assuming that similar results may be found 
for different flavor numbers, we have made a conjecture about the 
QCD phase diagram which, even if rather speculative, can be carefully checked by
future studies at imaginary $\mu_B$.
In particular one should clarify if a phase structure
like that sketched in Fig.~\ref{figdiag} is really valid, at least 
for $\mu^2 < 0$.

\section*{Acknowledgments}

We thank C.~Bonati, G.~Cossu, A.~Di Giacomo, Ph.~de Forcrand, A.~Patella, C.~Pica and E.~Vicari for 
useful discussions. M. D'E. thanks the organizers of the 
Workshop "Quarks, Hadrons, and the Phase Diagram of QCD" in St. Goar, where 
the present work has been completed.

\end{document}